# Self-similarity and scaling of thermal shock fractures

S. Tarasovs, A. Ghassemi

Department of Petroleum Engineering, Texas A&M University, College Station, TX 77843, USA

**Abstract**

The problem of crack pattern formation due to thermal shock loading at the surface of half-space is solved numerically using two-dimensional boundary element method. The results of numerical simulations with 100-200 random simultaneously growing and interacting cracks are used to obtain scaling relations for crack length and spacing. The numerical results predict that such process of pattern formation with quasi-static crack growth is not stable and at some point the excess energy leads to unstable propagation of one of the longest crack. The onset of instability has also been determined from numerical results.

Development of a hierarchical crack patterns is common in failure of brittle material in response to loading by a thermal shock. The thermally induced stresses are released by formation of an initial array of small cracks that grow in time as the cooling front propagates into the body, forming a system of cracks of different lengths. A similar process is the development of dessication cracks in mud/paste drying [1-3] or columnar joint formation in cooling lava lakes [4-6]. Chemical decomposition of solids also can generate crack patterns [7, 8]. Depending on the cooling/drying conditions, different crack patterns can be formed [9-14]. Many important characteristics of the structures such as fluid and heat transport properties depend on the number and length of the cracks, therefore significant efforts were undertaken to develop the theory of thermal shock fracturing. In [15, 16] the combination of strength theory and fracture mechanics was used to study the initiation and propagation of cracks due to thermal shock of brittle solid. The development of hierarchical crack patterns was explained in [17, 18] by bifurcation instability analysis. In particular, it was concluded that at a certain length, the quasi-static propagation of an array of equidistant cracks becomes unstable and only every second crack continues to grow until a new instability point, where a reduced number of cracks would continue to propagate, is reached. The formation of crack pattern in quenched glass/ceramic slabs was studied both experimentally and theoretically in [19-22].

Most of the existing studies of the hierarchical crack pattern formation used simplified model of symmetric equidistant edge cracks. Real materials are not homogeneous and the locations of cracks at the moment of initiation are affected by the local variation of the material's strength, so that the generated crack pattern is not symmetric. However, it can be expected that in average sense, the crack pattern which develops from such random array of cracks has deterministic characteristics [23]. Only few works have studied the formation of such random crack patterns. In [24] the crack pattern formation due to thermal shock loading was modelled using a simplified potential for crack growth and interaction, and it was found that the average crack spacing does not depend on the initial crack configuration. Similar results were obtained in [25] using complex hypersingular integral equation (CHIE) [26] method. In this paper, the CHIE method is used to simulate the simultaneous growth of many random cracks, and to study the scaling laws that govern the formation of crack patterns resulting from instantaneous cooling of the surface of a half-space.

Consider a half-space with an initial temperature $T_0$, subjected to instant cooling at its surface using a temperature drop of $\Delta T = T_0 - T_S$, where $T_S$ is the half-space surface temperature. The problem can be solved analytically, and the temperature profile at any time moment equals [27]

$$T(z) = T_0 - \Delta T \operatorname{erfc}\left(\frac{z}{L}\right), \tag{1}$$

where z is the distance from the surface, $L$ is the cooling depth which equals $L = \sqrt{4t\kappa}$, $t$ is time, and $\kappa$ is the thermal diffusivity of the solid. The cooling of the surface creates a thermally induced stresses in the material, and for two-dimensional plane strain condition, the tangential thermal stress component is given by:

$$\sigma_{th}(z) = \frac{E\alpha(T_0 - T(z))}{1-\nu}, \tag{2}$$

where $E$ is Young's modulus, $\nu$ is Poisson's ratio, and α is coefficient of linear thermal expansion.

The process of crack growth due to the thermal shock loading has two intrinsic length scales: the depth of the cooling zone $L$, and the characteristic length of the material ξ, defined as

$$\xi \equiv \left(\frac{K_{Ic}(1-\nu)}{E\alpha\Delta T}\right)^2, \tag{3}$$

with $K_{Ic}$ being fracture toughness of the material. The characteristic length ξ is the ratio of the energy required to create new crack surface and the thermoelastic energy that is generated in the solid by the thermal shock. From dimensional considerations the stress intensity factor (SIF) at the tip of a single edge crack of length $a$, normal to the solid surface and loaded by thermally induced stress can be expressed as:

$$K_I = \frac{E}{1-\nu}\alpha\Delta T \frac{L}{\sqrt{a}} \times f\left(\frac{a}{L}\right), \tag{4}$$

where the non-dimensional function $f$ has to be determined numerically and can be approximated as:

$$f\left(\frac{a}{L}\right) = 0.87 \tanh\left(2.2\frac{a}{L}\right). \tag{5}$$

For short cracks, i.e., when $a/L<0.5$, the SIF is approximately proportional to $\sqrt{a}$ and such crack is unstable. For long cracks (when $a/L>1$) the function $f$ is approximately constant and the crack length can be numerically estimated as:

$$a = \left(0.87 \frac{E\alpha\Delta T}{K_{Ic}(1-\nu)}\right)^2 L^2 . \qquad (6)$$

Therefore, as $L = \sqrt{4t\kappa}$, the length of the single crack subjected to thermal shock is a linear function of time.

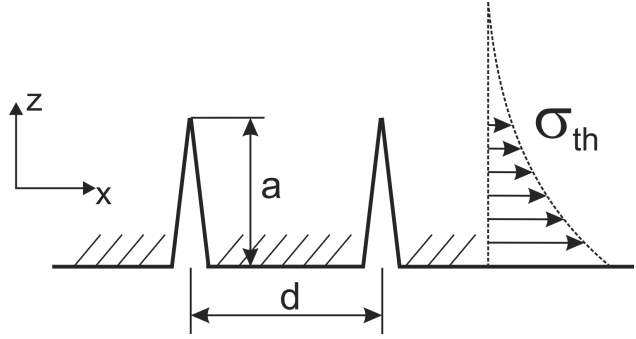

Fig.1: Array of edge cracks with length $a$ and spacing $d$, subjected to the thermally induced stresses $\sigma_{th}$.

To study the process of interaction of many cracks, the two-dimensional boundary element method was used. The simulations start from an initial array of many small cracks with length $a$ and average spacing $d$, as shown in Fig. 1. Using the superposition principle, the thermal load is applied at the faces of the cracks. The randomness of the initial crack array is introduced via perturbations in the cracks locations. Each initial crack is shifted from its position by a random value within $\pm\Delta d$ while keeping the average spacing between the cracks, $d$, constant. Three different levels of randomness were tested, with $\Delta d/d$ equal 0.1, 0.2 and 0.4. The simulation results show that after several crack increments, when some of the initial cracks stop, the resulting crack pattern does not depend on the initial configuration in the average sense so that, the crack spacing for different parameters fall into single master curve irrespective of the initial cracks configuration. For further simulations only $\Delta d/d$ value of 0.2 was used. To replicate a large number of cracks, an initial array of 100-200 small cracks with periodic boundary conditions was used (so that the whole random array is repeated) in the simulations. For accurate determination of the average crack spacing, normally about 6 independent simulations with different random crack locations were performed for each value of characteristic length $\xi$.

The final crack patterns for different values of characteristic length are presented in Fig. 2. The numerical results suggest that the process of crack pattern formation is self-similar, i.e., the crack pattern repeats itself on different time and length scales. The results of simulations, i.e., the maximum crack length and average crack spacing normalized with respect to material constant $\xi$, are presented in Figs. 3 and 4 in logarithmic scale. The power law curve fits shown correspond to Eqs. (7) and (8), respectively:

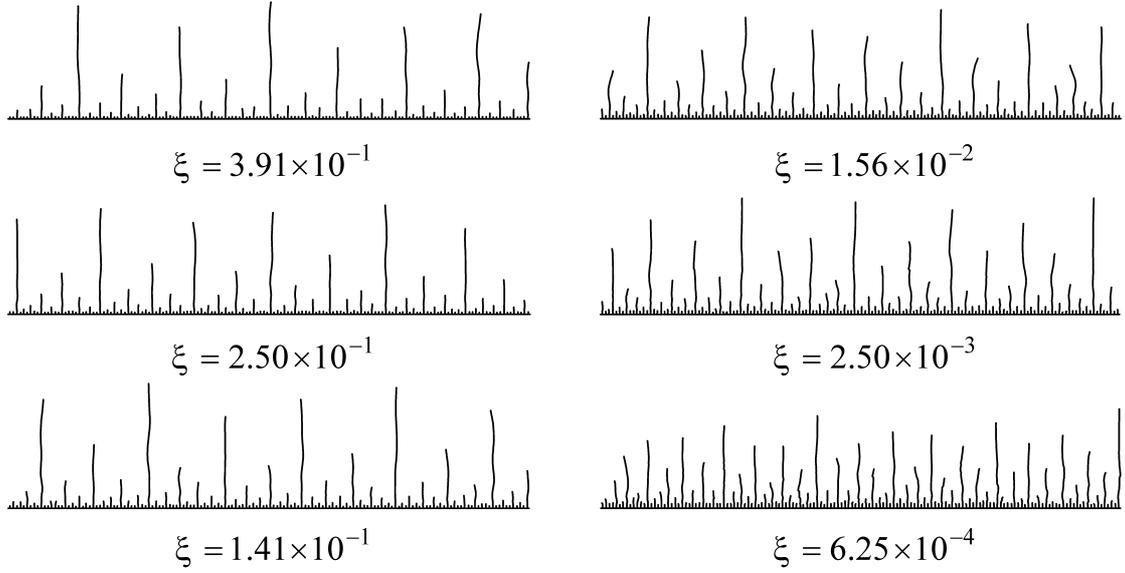

Fig. 2: Crack patterns formed in simulations of thermal shock fracture with different values of characteristic length, $\xi$. The time and length scales are different in each figure but the pattern development in time is similar.

$$\frac{a}{\xi} = \left(\frac{L}{\xi}\right)^{1.09}, \tag{7}$$

$$\frac{d}{\xi} = 4.4\left(\frac{z}{\xi}\right)^{0.77}. \tag{8}$$

From Eq. (7) it follows that the depth of the random array of thermal cracks is approximately proportional to the square root of time. It should be noted that the scaling laws (7) and (8) are quite close to the scaling derived in [28] using simplified bifurcation analysis. In [28] simple relation between crack length and spacing was obtained as $ad = 1.74L^2$. Combining Eqs. (7) and (8) yields $ad = 4.4\xi^{0.07}L^{1.93}$. The scaling relation for crack spacing in [28] has different form than Eq. (8), but the numerical values for the normalized depth in the range of $10^1 - 10^4$ are quite close. Outside of this range, the solution is not physically meaningful. For the $a/\xi < 10$, a crack initiation criterion has to be applied to determine the smallest possible crack size and spacing. The analysis of [29, 23] shows that the initial crack length is of the order of $\xi$. Only after the initial cracks advance and begin to interact, do the scaling laws (7) and (8) can be applied.

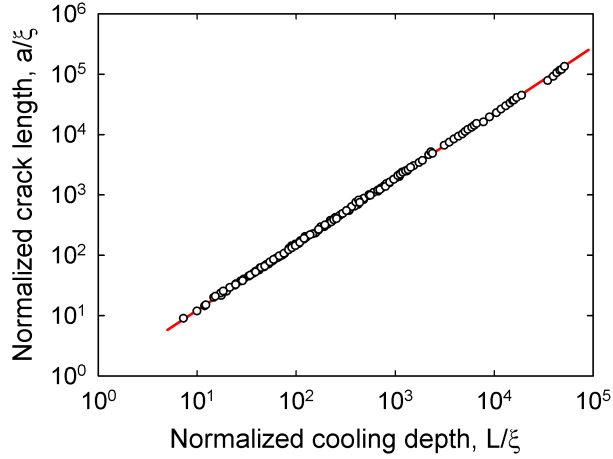

Fig. 3: Scaling relation between crack length *a* and cooling depth *L*. Symbols – numerical results, line – power law fit (Eq. (7)).

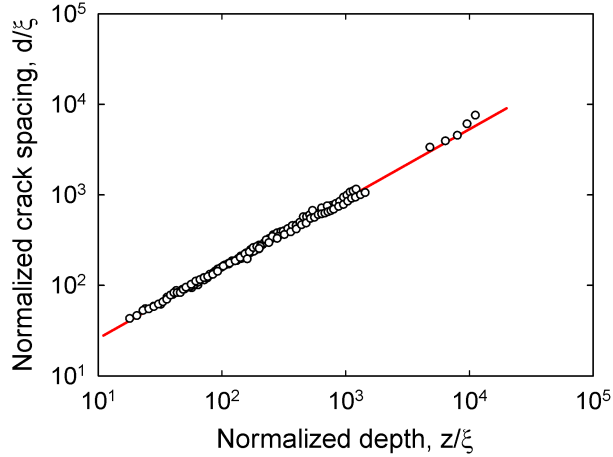

Fig. 4: Scaling relation between crack spacing *d* and depth. Symbols – average numerical results, line – power law fit (Eq. (8)).

For the long range, $a/\xi > 10^4$, the solution becomes unstable. The elastic energy induced in the body by cooling is proportional to the cooling depth *L*. Using relations (7) and (8), the total length of all cracks for a unit length of the cooled surface can be estimated as:

$$S_{total} = \int_0^a \frac{1}{d} dz \sim \xi^{-0.25} L^{0.25}. \qquad (9)$$

Eq. (9) predicts that the total length of all cracks grows much slower than the elastic energy. Excess energy is accumulated in the system, and this energy eventually is released by unstable growth of some cracks. Physically, this means that at some stage of propagation, the classical alternating bifurcation solution [17, 18], where every second crack stops at the bifurcation point, is no longer favorable and it is replaced by another bifurcation solution with only a single growing crack. We have observed such process in our numerical simulations when the length of the cracks was sufficiently large.

In Eq. (9) we have assumed that an infinite number of infinitesimal cracks exist. Since both in real materials and numerical simulations the process of thermal shock cracking starts from an initial array of cracks with finite length, the onset of instability depends on the initial configuration. However, if initial cracks are sufficiently small (of the order of $\xi$), the onset of instability predicted by numerical simulations can be approximated by power law function $a_{cr} = C\xi^\lambda$, where C is a dimensional parameter. The critical crack length is plotted in Fig. 5 as a function of characteristic length together with the power law fit: $a_{cr} = 520\xi^{0.72}$. The time of for the onset of instability can be estimated as $L_{cr} = 280\xi^{0.75}$.

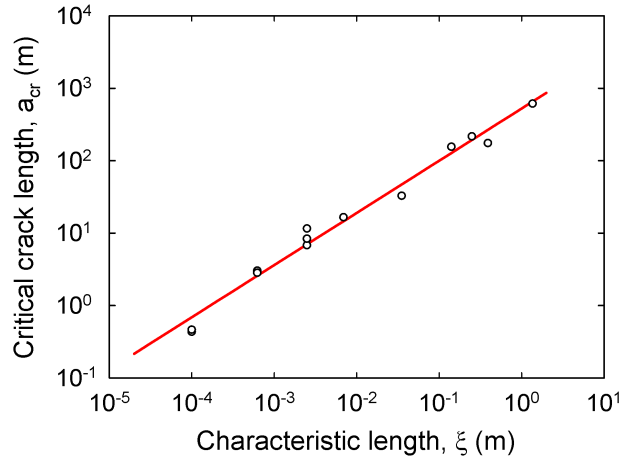

Fig. 5: Onset of instability during quasi-static growth of array of edge cracks. Symbols – numerical results, line – power law fit.

In conclusion, extensive two-dimensional numerical simulations of the thermal shock cracking has been performed using the complex variable hypersingular boundary element method with a periodic array of about 100-200 simultaneously growing random cracks. The numerical results have shown that the crack pattern is self-similar, and the scaling relations for crack length and crack spacing were obtained by analyzing the numerically simulated patterns. It is found that the total length of all cracks grows much slower than the strain energy of the thermal stress due to cooling. This excess energy may lead to unstable propagation of some cracks. Such process has been observed in numerical simulations and has been used to determine the onset of instability.

Acknowledgements

This project was supported by the U.S. Department of Energy Office of Energy Efficiency and Renewable Energy under Cooperative Agreement DE-PS36-08GO1896. This support does not constitute an endorsement by the U.S. Department of Energy of the views expressed in this publication. Also, S. Tarasovs gratefully acknowledges Dr. J. Andersons of Institute of Polymer Mechanics, Latvia, for helpful discussions.